\documentclass[aps,twocolumn]{revtex4}
\usepackage[colorlinks, linkcolor=red,anchorcolor=green,citecolor=blue]{hyperref}
\usepackage{graphicx}
\usepackage{bm}
\usepackage{amsmath}
\usepackage{color}
\usepackage{xcolor, color, framed, ulem}
\graphicspath{{./fig/}}

\begin{document}

\title{Mass spectra of doubly heavy tetraquarks in an improved chromomagnetic interaction model}
\author{Tao Guo}
\author{Jianing Li}
\author{Jiaxing Zhao}
\author{Lianyi He}
\affiliation{Physics Department, Tsinghua University, Beijing 100084, China\\}

\date{\today}%

\begin{abstract}
Doubly heavy tetraquark  states are the prime candidates for tightly bound exotic states. We present a systematic study of the mass spectra of the $S$-wave doubly heavy tetraquark states $QQ\bar{q}\bar{q}$ ($q=u, d, s$ and $Q=c, b$) with different quantum numbers $J^P=0^+$, $1^+$, and $2^+$ in the framework of the improved chromomagnetic interaction (ICMI) model.
The parameters in the ICMI model are obtained by fitting the conventional hadron spectra and are used directly to predict the masses of the tetraquark states. For heavy quarks, the uncertainties of the parameters are obtained by comparing the masses of doubly and triply heavy baryons with those given by lattice QCD, QCD sum rules, and potential models. Several compact and stable bound states are found in both the doubly charmed and doubly bottomed tetraquark systems. The predicted mass of the $cc\bar u\bar d$ state is consistent with the recent measurement from the LHCb collaboration.
\end{abstract}


\maketitle

\section{Introduction}
Quantum chromodynamics (QCD) allows the existence of many fantastic hadrons, such as glueballs~\cite{Mathieu:2008me,Celi:2013gma}, hybrids~\cite{Meyer:2015eta,Chanowitz:1982qj}, multi-quark states (e.g. tetraquarks and pentaquarks)~\cite{Esposito:2016noz,Karliner:2017qhf}, and hadronic molecules~\cite{DeRujula:1976zlg,Guo:2017jvc}. The flavor SU(4) quark model predicts 22 charmed baryons~\cite{GellMann:1964nj,Zyla:2020zbs}, but many of them haven't yet been discovered, such as the triply charmed baryon $\Omega_{ccc}$ and the doubly charmed baryons $\Xi^+_{cc}(ccd)$, $\Xi^{++}_{cc}(ccu)$, and $\Omega_{cc}(ccs)$.
Experimentalists have been searching for these doubly and triply charmed baryons for decades. In 2003, the SELEX collaboration claimed the discovery of $\Xi_{cc}^+(ccd)$ in the decay channel $\Xi_{cc}^+\to\Lambda_c^+ K^-\pi^+$~\cite{SELEX:2002wqn} , however, the Belle, BaBar, and LHCb collaborations failed to reproduce their results after that~\cite{Belle:2006edu,BaBar:2006bab,LHCb:2013hvt}.
In 2017, the LHCb Collaboration claimed the discovery of $\Xi_{cc}^{++}(ccu)$ in the $\Lambda_c^+ K^-\pi^+\pi^+$ mass spectrum~\cite{LHCb:2017iph}. The $\Xi_{cc}^{++}$ mass was determined to be 3621.40 MeV, which agrees with previous theoretical predictions~\cite{Roncaglia:1995az,Brown:2014ena,PACS-CS:2013vie,Kiselev:2001fw,Shi:2019tji,Martynenko:2007je,Gershtein:2000nx,Mehen:2006vv}. The production cross sections of these doubly charmed baryons in proton-proton~\cite{Berezhnoy:1998aa} and heavy-ion collisions~\cite{Zhao:2016ccp} were also investigated, and it was predicted that the cross sections could be greatly enhanced in heavy-ion collisions~\cite{Zhao:2016ccp}.
Moreover, the discoveries of the exotic XYZ mesons~\cite{Brambilla:2019esw} and the fully charmed tetraquark state $X(cc\bar c\bar c)$~\cite{LHCb:2020bwg} have attracted much attention.

The observations of the doubly charmed baryon $\Xi_{cc}^{++}(ccu)$ and the fully charmed tetraquark state $X(cc\bar c\bar c)$ indicate that if the doubly charmed tetraquark state $(cc\bar u\bar d)$ exists, it should be accessible in the $DD^*$ final states. Exhilaratingly, the LHCb Collaboration have recently found a very narrow peak near the $DD^*$ threshold in the $D^0D^0\pi^+$ invariant mass spectrum~\cite{LHCb:Polyakov-1,LHCb:Polyakov-2}. The preliminary results show that the quantum number of this state is $J^P=1^+$ and the mass is only $273$ KeV lower than the $D^0D^{*+}$ threshold.
On the theoretical side, the doubly charmed tetraquark states have been anticipated for more than 40 years. They are the prime candidates of tightly bound exotic states that decay only weakly. The mass spectra and possible decay channels of the doubly charmed tetraquark states have been studied extensively by using the quark models~\cite{Karliner:2017qjm,Eichten:2017ffp,Silvestre-Brac:1993zem,Lu:2020rog,Luo:2017eub,Park:2018wjk,Ebert:2007rn,Yang:2009zzp,Yan:2018gik,Gelman:2002wf,Cheng:2020wxa,Faustov:2021hjs,Hernandez:2019eox,Richard:2018yrm,Vijande:2007rf,Vijande:2009kj}, QCD sum rules~\cite{Navarra:2007yw,Dias:2011mi,Du:2012wp,Tang:2019nwv,Agaev:2021vur,Agaev:2019qqn,Wang:2017dtg}, effective field theory~\cite{Ding:2009vj,Ohkoda:2012hv,Xu:2017tsr,Qin:2020zlg}, and lattice QCD~\cite{Brown:2012tm,Ikeda:2013vwa,Leskovec:2019ioa,Bicudo:2015vta,Bicudo:2016ooe,Francis:2016hui,Junnarkar:2018twb,Francis:2018jyb}. Almost all these theoretical studies found that the masses of the doubly bottomed tetraquark states $(bb\bar q\bar q)$ are below the meson-meson thresholds, making them stable, but the masses of the doubly charmed tetraquark states $(cc\bar q\bar q)$ are above the thresholds. For the doubly heavy tetraquark states $bc\bar q\bar q$, some studies~\cite{Karliner:2017qjm,Weng:2021hje} suggest that their masses are somewhat below the 
$\bar DB$ threshold, while others predict that their masses are above the threshold~\cite{Eichten:2017ffp,Luo:2017eub,Lu:2020rog,Faustov:2021hjs}.  The improved chromomagnetic interaction model (ICMI)~\cite{Hogaasen:2013nca,Weng:2018mmf,An:2021vwi}, which incorporates both chromomagnetic and chromoelectric interaction effects, has never been used to study the doubly heavy tetraquark states. In this work, we employ this model to study the doubly heavy tetraquark systems and search for new stable bound states.

This paper is organized as follows. In Section II, we give a brief introduction of the ICMI model and discuss its application to the doubly heavy tetraquark systems.  In Section III, we determine the parameters in the ICMI model, such as effective masses and coupling strengths. The mass spectra of various doubly heavy tetraquark states are calculated and presented in this section. We summarize in Section IV.

\section{ Theoretical approach}
The quark model assumes that hadrons are color singlet bound states of constituent quarks. The interactions between two constituent quarks can be given simply by the one-gluon-exchange (OGE) process~\cite{DeRujula:1975qlm},
\begin{equation}
V_{ij}^{\text{OGE}}={\alpha_s\over 4}{\lambda}_{i}^{c}\cdot{\lambda}_{j}^{c}\left({1\over r_{ij}}-{2\pi \delta({\bf r}_{ij})\mbox{\boldmath{$\sigma$}}_i\cdot\mbox{\boldmath{$\sigma$}}_j\over 3m_im_j} \right),
\end{equation}
where $m_i$ is the effective mass of the $i$-th constituent quark, $\alpha_s$ is the running coupling constant, $r_{ij}=|{\bf r}_{ij}|=|{\bf r}_i-{\bf r}_j|$ is the distance between the $i$-th and $j$-th quarks. ${\lambda}_{i}^{c}$ ($c=1, ..., 8$) are the Gell-Mann matrices acting on the color space of the $i$-th quark, and $\mbox{\boldmath{$\sigma$}}_i$ are the Pauli matrices on the spin space of the $i$-th quark.
The chromomagnetic interaction corresponds to the spin-dependent part and is given by
\begin{equation}\label{cmv}
V^{\text{cm}}_{ij}=-{\alpha_s\pi \delta({\bf r}_{ij})\over 6m_im_j}{\lambda}_{i}^{c}\cdot{\lambda}_{j}^{c}\mbox{\boldmath{$\sigma$}}_i\cdot\mbox{\boldmath{$\sigma$}}_j,
\end{equation}
while the chromoelectric interaction reads
\begin{equation}\label{cev}
V^{\text{ce}}_{ij}={\alpha_s\over 4r_{ij}}{\lambda}_{i}^{c}\cdot{\lambda}_{j}^{c}.
\end{equation}
If we focus on the $S$-wave states, the spin-orbit interactions can be neglected. The improved chromomagnetic interaction model (ICMI) can be obtained by integrating over the spatial wave function. The Hamiltonian of a four-body system composed of four quarks in the ICMI model can be expressed as~\cite{Hogaasen:2013nca,Weng:2018mmf,An:2021vwi,DeRujula:1975qlm,Liu:2019zoy,Maiani:2004vq,Cui:2005az,Hogaasen:2005jv,Guo:2011gu,Kim:2014ywa}:
\begin{equation}\label{Ha}
H=\sum\limits_{i=1}^4m_i+H_{\text{cm}} + H_{\text{ce}},
\end{equation}
where the chromomagnetic interaction term is given by
 \begin{equation}\label{cm}
H_{\text{cm}}=-\sum\limits_{i<j}v_{ij}{\lambda}_{i}^{c}\cdot{\lambda}_{j}^{c}
\mbox{\boldmath{$\sigma$}}_i\cdot\mbox{\boldmath{$\sigma$}}_j,
\end{equation}
and the chromoelectric interaction term reads
\begin{equation}\label{ce}
H_{\text{ce}}=-\sum\limits_{i<j}c_{ij}{\lambda}_{i}^{c}\cdot{\lambda}_{j}^{c}.
\end{equation}
Furthermore, if the subscript $i$ (or $j$) denotes an antiquark, $\lambda_i^c$ should be replaced by $-\lambda_i^{c*}$. The parameters $v_{ij}$ and $c_{ij}$ are formally derived by integrating over the spatial wave function. They incorporate the effects of the spatial configuration, effective quark masses, and the coupling constant. In the ICMI model, they are regarded as model parameters that are usually obtained by fitting the known hadron spectra. 
Introducing the parameter $m_{ij}=m_i+m_j+16c_{ij}/3$~\cite{Weng:2018mmf,Weng:2019ynv}, we can rewrite the Hamiltonian as
 \begin{equation}\label{Ha1}
 H=H_0+H_{\text{cm}},
 \end{equation}
where
\begin{equation}\label{H0}
H_0=-\frac{3}{16}\sum\limits_{i<j}m_{ij}{\lambda}_{i}^{c}\cdot{\lambda}_{j}^{c}.
 \end{equation}
We note that the parameter $m_{ij}$ is related to the effective masses of the constituent quarks and the coupling strength $c_{ij}$ of the chromoelectric interaction.

For the doubly heavy tetraquark state $Q_1Q_2 \bar{q}_3\bar{q}_4$, where $Q$ denotes a heavy quark and $q$ a light quark ($u$ or $d$ quark) or a strange quark, there are two types of decomposition of the wave function in color space based on the SU$(3)$ group theory. They physically correspond to two different configurations in color space, the diquark-anti-diquark configuration labeled as $|(Q_1Q_2)(\bar q_3\bar{q}_4)\rangle$ and the meson-meson configuration labeled as $|(Q_1\bar{q}_3)(Q_2\bar{q}_4)\rangle$(or $|(Q_1\bar{q}_4)(Q_2\bar{q}_3)\rangle$). In each decomposition, there are two color-singlet states. For the 
meson-meson configuration, they are given by
\begin{eqnarray}
|(Q_1\bar q_3)^1(Q_2\bar q_4)^1\rangle^1, ~ |(Q_1\bar q_3)^8(Q_2\bar q_4)^8\rangle^1,
\end{eqnarray}
or the equivalent states with the permutation $1\leftrightarrow2$ or $3\leftrightarrow4$.  For the diquark-anti-diquark configuration, they are given by
\begin{eqnarray}
|(Q_1Q_2)^{\bar 3}(\bar q_3\bar q_4)^{3}\rangle^1, ~|(Q_1Q_2)^6(\bar q_3\bar q_4)^{\bar 6}\rangle^1.
\end{eqnarray}
The superscripts here denote the color channels of the subsystems $Q_1\bar q_3$ and $Q_2\bar q_4$ (or $Q_1Q_2$ and $\bar q_3\bar q_4$) and the whole tetraquark system.
These two sets of color-singlet basis are connected with each other through a unitary transformation. The corresponding matrix elements of the transformation can be obtained by
 acting the Casimir operator on the above color-singlet states.
Since the Pauli exclusion principle requires that the wave function should be antisymmetric,  it is convenient to use the diquark-anti-diquark configuration, i.e., the basis 
$|(Q_1Q_2)(\bar q_3\bar{q}_4)\rangle$.

Meanwhile, the direct product decomposition in spin space shows that the total spin of all possible $S$-wave tetraquark states can be 0, 1, and 2.  Using the diquark-anti-diquark configuration, we can construct the spin states as follows. There are two spin-zero states,
\begin{eqnarray}
&&|(Q_1Q_2)_0(\bar q_3\bar q_4)_0\rangle_0,\nonumber\\
&&|(Q_1Q_2)_1(\bar q_3\bar q_4)_1\rangle_0,
\end{eqnarray}
three spin-one states,
\begin{eqnarray}
&&|(Q_1Q_2)_0(\bar q_3\bar q_4)_1\rangle_1,\nonumber\\
&&|(Q_1Q_2)_1(\bar q_3\bar q_4)_0\rangle_1, \nonumber\\
&&|(Q_1Q_2)_1(\bar q_3\bar q_4)_1\rangle_1,
\end{eqnarray}
and one spin-two state,
\begin{eqnarray}
&&|(Q_1Q_2)_1(\bar q_2\bar q_4)_1\rangle_2,
\end{eqnarray}
where the subscripts denote the spin channels of the subsystems $Q_1Q_2$ and $\bar q_3\bar q_4$ and the whole tetraquark system.

Now, we can construct all possible basis wave functions in the color and spin spaces for the tetraquark systems with given quantum numbers $J^P=0^{+}$, $1^{+}$, and $2^{+}$. For the scalar tetraquark states with $J^P=0^{+}$, the color-spin basis can be built as
\begin{eqnarray}
 &&|(Q_1Q_2)^{\bar 3}_0\otimes(\bar q_3\bar{q}_4)^3_0\rangle \delta_{12}^S\delta_{34}^S,\nonumber\\
 &&|(Q_1Q_2)^{\bar 3}_1\otimes(\bar q_3\bar{q}_4)^3_1\rangle \delta_{12}^A\delta_{34}^A,\nonumber\\
 &&|(Q_1Q_2)^6_0\otimes(\bar q_3\bar{q}_4)^{\bar 6}_0\rangle \delta_{12}^A\delta_{34}^A, \nonumber\\
 &&|(Q_1Q_2)^6_1\otimes(\bar q_3\bar{q}_4)^{\bar 6}_1\rangle \delta_{12}^S\delta_{34}^S,
\end{eqnarray}
where the superscripts and subscripts again denote the color and spin channels of the subsystems $Q_1Q_2$ and $\bar q_3\bar{q}_4$, respectively.
For the axial vector tetraquark states with quantum number $J^P=1^+$, the color-spin basis is given by
\begin{eqnarray}\label{1324basis1}
&& |(Q_1Q_2)^{\bar 3}_0\otimes(\bar q_3\bar{q}_4)^3_1\rangle \delta_{12}^S\delta_{34}^A, \nonumber\\
&&|(Q_1Q_2)^{\bar 3}_1\otimes(\bar q_3\bar{q}_4)^3_0\rangle \delta_{12}^A \delta_{34}^S,\nonumber\\
&&|(Q_1Q_2)^{\bar 3}_1\otimes(\bar q_3\bar{q}_4)^3_1\rangle \delta_{12}^A \delta_{34}^A,\nonumber\\
&&|(Q_1Q_2)^6_0\otimes(\bar q_3\bar{q}_4)^{\bar 6}_1\rangle \delta_{12}^A \delta_{34}^S,\nonumber\\
&&|(Q_1Q_2)^6_1\otimes(\bar q_3\bar{q}_4)^{\bar 6}_0\rangle \delta_{12}^S \delta_{34}^A,\nonumber\\
&&|(Q_1Q_2)^6_1\otimes(\bar q_3\bar{q}_4)^{\bar 6}_1\rangle \delta_{12}^S \delta_{34}^S.
\end{eqnarray}
For the $J^{P}=2^{+}$ states, the color-spin basis reads
\begin{eqnarray}\label{1324basis2}
&&|(Q_1Q_2)^{\bar 3}_1\otimes(\bar q_3\bar{q}_4)^3_1\rangle \delta_{12}^A \delta_{34}^A, \nonumber\\
&&|(Q_1Q_2)^6_1\otimes(\bar q_3\bar{q}_4)^{\bar 6}_1\rangle \delta_{12}^S \delta_{34}^S.
\end{eqnarray}

In the above basis wave functions, the symbol $\delta_{ij}^S$ and $\delta_{ij}^A$ are introduced to ensure the Pauli exclusion principle, i.e.,  the exchange symmetry between the $i$-th and $j$-th quarks or antiquarks. If $Q_1$ and $Q_2$ are the same heavy flavor, the flavor wave function is symmetric. Because the total wave function of the $Q_1Q_2$ sector should be antisymmetric, we need to set $\delta_{12}^S=0$ and $\delta_{12}^A=1$. If $Q_1$ and $Q_2$ are different heavy flavors, we simply set $\delta_{12}^S=\delta_{12}^A=1$. On the other hand, 
if $\bar q_3$ and $\bar q_4$ are both light antiquarks (anti-$u$ and -$d$ quarks), the symmetry of flavor wave function of the $\bar q_3\bar q_4$ sector depends on the isospin. For the isospin singlet state, 
we set
$\delta_{34}^S=1$ and $\delta_{34}^A=0$. For the isospin triplet state, we set $\delta_{34}^S=0$ and $\delta_{34}^A=1$. Finally, if $\bar q_3$ and $\bar q_4$ are a pair of a light antiquark and a strange antiquark, we have $\delta_{34}^S=\delta_{34}^A=1$.

Having constructed a representation based on the above basis wave functions,  we can obtain the matrix form of the Hamiltonian (\ref{Ha1}).
The mass spectra of all possible tetraquark states can be obtained by diagonalizing this matrix. Finally, when we analyze the decay channels of the tetraquark states,  it is convenient to convert the  diquark-anti-diquark configuration to the meson-meson configuration via a transformation of representation.

\section{Results and discussion}
\subsection{Model parameters}

\begin{table}[!htbp]
\caption{\label{mesonpm} Parameters $m_{ij}$ and $v_{ij}$ (in MeV) for quark-antiquark systems.}
\begin{ruledtabular}
\begin{tabular}{lcccc}
$m_{ij}$ & $n$ & $s$ & $c$ & $b$\\
\hline
$\bar{n}$&616.34&&&\\
$\bar{s}$&792.17&$963.43$&&\\
$\bar{c}$&1975.11&2076.24&3068.67&\\
$\bar{b}$&5313.36&5403.28&$6327.40$&$9444.91$\\
\hline\hline
$v_{ij}$ & $n$ & $s$ & $c$ & $b$\\
 \hline
$\bar{n}$&29.798&&&\\
$\bar{s}$&18.656&$10.506$&&\\
$\bar{c}$&6.591&6.743&5.298&\\
$\bar{b}$&2.126&2.273&$3.281$&$2.888$\\
\end{tabular}
\end{ruledtabular}
\end{table}

\begin{table}[!htbp]
\caption{\label{baryonpm} Parameters $m_{ij}$ and $v_{ij}$ (in MeV) for quark-quark (antiquark-antiquark) systems.}
\begin{ruledtabular}
\begin{tabular}{lcccc}
$m_{ij}$ & $n$ & $s$ & $c$ & $b$\\
\hline
$n$&723.86& & &\\
$s$&904.83&1080.59& &\\
$c$&2085.58&2185.99&3182.67$\pm30$ &\\
$b$&5413.07&5510.83&6441.40$\pm30$ &9558.91$\pm60$\\
\hline\hline
$v_{ij}$ & $n$ & $s$ & $c$ & $b$\\
\hline
$n$&18.277&&&\\
$s$&12.824&6.445& &\\
$c$&4.063&4.148&3.354$\pm0.5$ & \\
$b$&1.235&1.304&2.077$\pm0.5$ &1.448$\pm0.5$ \\
\end{tabular}
\end{ruledtabular}
\end{table}

To calculate the mass spectra of the doubly heavy tetraquark states $QQ\bar{q}\bar{q}$ within the ICMI model, we first need to determine the model parameters $m_{ij}$ and $v_{ij}$. Most of them can be obtained by fitting the the masses of the ground-state mesons and baryons determined in previous experiments~\cite{Zyla:2020zbs}.
The parameters for the quark-antiquark systems has been be extracted~\cite{Guo:2021mja} and are displayed in Table~\ref{mesonpm}. Note that the symbol $n$ is used to denote the light quarks ($u$ and $d$ quarks) in the following.
Since the $s\bar{s}$ meson state with $J^P=0^-$ and the $b\bar{c}$ meson state with $J^P=1^-$ haven't yet been observed experimentally, the parameters $v_{s\bar{s}}$ and $v_{c\bar{b}}$ are obtained via a similar strategy as shown in Ref.~\cite{Wu:2018xdi}.
It is worth noting that the parameter matrices are symmetric, $v_{ij}=v_{ji}$ and $m_{ij}= m_{ji}$. Therefore, only half of the off-diagonal elements are listed in Table~\ref{mesonpm}.

\begin{table*}[!htbp]
\caption{Mass spectra of doubly and triply heavy baryons (in GeV) from lattice QCD, QCD sum rules, potential models, and the ICMI model (this work).\label{table3}}
\vspace{0.2cm}
\centering
\begin{tabular}{llcccccc}
\hline
\hline
 States & $J^P$ &  Lattice~\cite{PACS-CS:2013vie,Lewis:2008fu,Meinel:2010pw} & Lattice~\cite{Brown:2014ena} & QCD sum rules~\cite{Kiselev:2001fw,Wang:2020avt} & Potential~\cite{Ebert:2002ig,Shi:2019tji,Zhao:2020jqu} & Potential~\cite{Martynenko:2007je} & This work\\
\hline
$\Xi_{cc}$          &1/2& 3.603 &3.610    &3.48 & 3.620  &3.510    &3.627$\sim$3.657\\
$\Xi_{cc}$          &3/2& 3.706 &3.692    &3.61 & 3.727  &3.548    &3.692$\sim$3.722\\
$\Omega_{ccc}$&3/2& 4.789 &4.796    &4.81  & 3.784  & 4.803   &4.756$\sim$4.846\\
$\Xi_{bb}$          &1/2& 10.127&10.143 &10.09&10.202 & 10.130 &10.153$\sim$10.213\\
$\Xi_{bb}$          &3/2& 10.151&10.178 &10.13&10.237 & 10.144 &10.173$\sim$10.233\\
$\Omega_{bbb}$&3/2& 14.371&14.366 &14.43      &14.499 & 14.569 &14.260$\sim$14.440\\
$\Omega_{bcc}$ &1/2& -&8.007 &8.02 &8.143 & 8.018 & 7.989$\sim$8.049\\
$\Omega_{bcc}$ &3/2& -&8.037 &8.03&8.207 & 8.025 & 8.022$\sim$8.082\\
$\Omega_{bbc}$ &1/2& -&11.195 &11.22&10.920 & 11.280 &11.172$\sim$11.232\\
$\Omega_{bbc}$ &3/2& -&11.229 &11.23&10.953 & 11.287 &11.205$\sim$11.265\\
\hline
\hline
\end{tabular}
\end{table*}

The parameters for the quark-quark systems can be extracted from known baryon mass spectra. Detailed analysis and procedure are demonstrated in Ref.~\cite{Guo:2021mja}. For light and strange quarks, the values of the model parameters $m_{qq}$ and $v_{qq}$ are displayed in Table~\ref{baryonpm}. Note that for the antiquark-antiquark systems, we have
$m_{\bar{q}\bar{q}}=m_{qq}$ and $v_{\bar{q}\bar{q}}=v_{qq}$ according to the charge conjugate symmetry. 
However, for heavy quark pairs, it is difficult to extract the parameters $m_{QQ}$ and $v_{QQ}$, not only for the lack of the experimental data on doubly and triply heavy baryons but also for the theoretical challenges. In the ICMI model, the masses of the $S$-wave baryons with two identical heavy quarks and total spin $S=1/2$ are given by
\begin{eqnarray}\label{Baryonmass}
M_{QQq}=\frac{1}{2}(m_{QQ}+2m_{Qq})+\frac{8}{3}(v_{QQ}-4v_{Qq}).
\end{eqnarray}
While for total spin $S = 3/2$, the masses read
\begin{equation}\label{Baryonmass2}
M_{QQq}=\frac{1}{2}(m_{QQ}+2m_{Qq})+\frac{8}{3}(v_{QQ}+2v_{Qq}).
\end{equation}
On the other hand, for triply heavy baryons with three identical heavy quarks, the total spin can only be $S=3/2$ and the mass is given by
\begin{equation}
M_{QQQ}=\frac{3}{2}m_{QQ}+8v_{QQ}.
\end{equation}
According to the above mass formulae, we find that it is impossible to extract the model parameters $m_{QQ}$ and $v_{QQ}$ independently from different doubly (or triply) heavy baryons. However, the combination of $m_{QQ}$ and $v_{QQ}$, $3m_{QQ}/2+8v_{QQ}$, is ascertainable and can be given directly by the mass of the $\Omega_{QQQ}$ baryon.  
Lattice QCD~\cite{PACS-CS:2013vie,Lewis:2008fu,Meinel:2010pw,Brown:2014ena}, QCD sum rules~\cite{Kiselev:2001fw,Wang:2020avt}, and potential models~\cite{Roncaglia:1995az,Ebert:2002ig,Shi:2019tji,Martynenko:2007je,Ebert:1996ec,Gershtein:2000nx,Tong:1999qs} predict the masses of $\Omega_{ccc}$ and $\Omega_{bbb}$ to be around 4800 MeV and 14000 MeV, respectively. If the coupling strength $v_{QQ}$ is determined, we can obtain the value of $m_{QQ}$ by using the mass of the $\Omega_{QQQ}$ baryon. In this work, we determine the parameter  $v_{QQ}$ via the relation $v_{QQ}/v_{Q\bar Q}=v_{qq}/v_{q\bar q}$~\cite{Wu:2016vtq}. We also take into account the uncertainties of the masses of triply heavy baryons and the coupling strength $v_{QQ}$,  which give rise to error bars for the parameters $m_{QQ}$ and $v_{QQ}$.  The central values and the errors of $m_{QQ}$ and $v_{QQ}$ are shown in Table~\ref{baryonpm}.
The parameters $m_{Qq}$ and $v_{Qq}$ can be determined via the mass formulae (\ref{Baryonmass}) and (\ref{Baryonmass2}) with known baryon mass spectra. For the parameters $m_{cb}$ and $v_{cb}$, we extract them from the mass spectra of triply heavy baryons (e.g. $\Omega_{ccb}$ and $\Omega_{bbc}$) with the same strategy. Using the parameters shown in Table~\ref{baryonpm}, we
recalculate the masses of the doubly and triply heavy baryons. The results are shown in Table~\ref{table3}. The baryon masses from the present ICMI model are in a range due to the consideration of the errors of the model parameters. We find that the baryon masses obtained from the present ICMI model are consistent with the results from other theoretical approaches.  

\subsection{Mass spectra of tetraquark states $QQ\bar q\bar q$}

\begin{figure*}[htbp]
\centering
\begin{minipage}{6.39cm}
    \includegraphics[width=0.96\textwidth,height=0.96\textwidth]{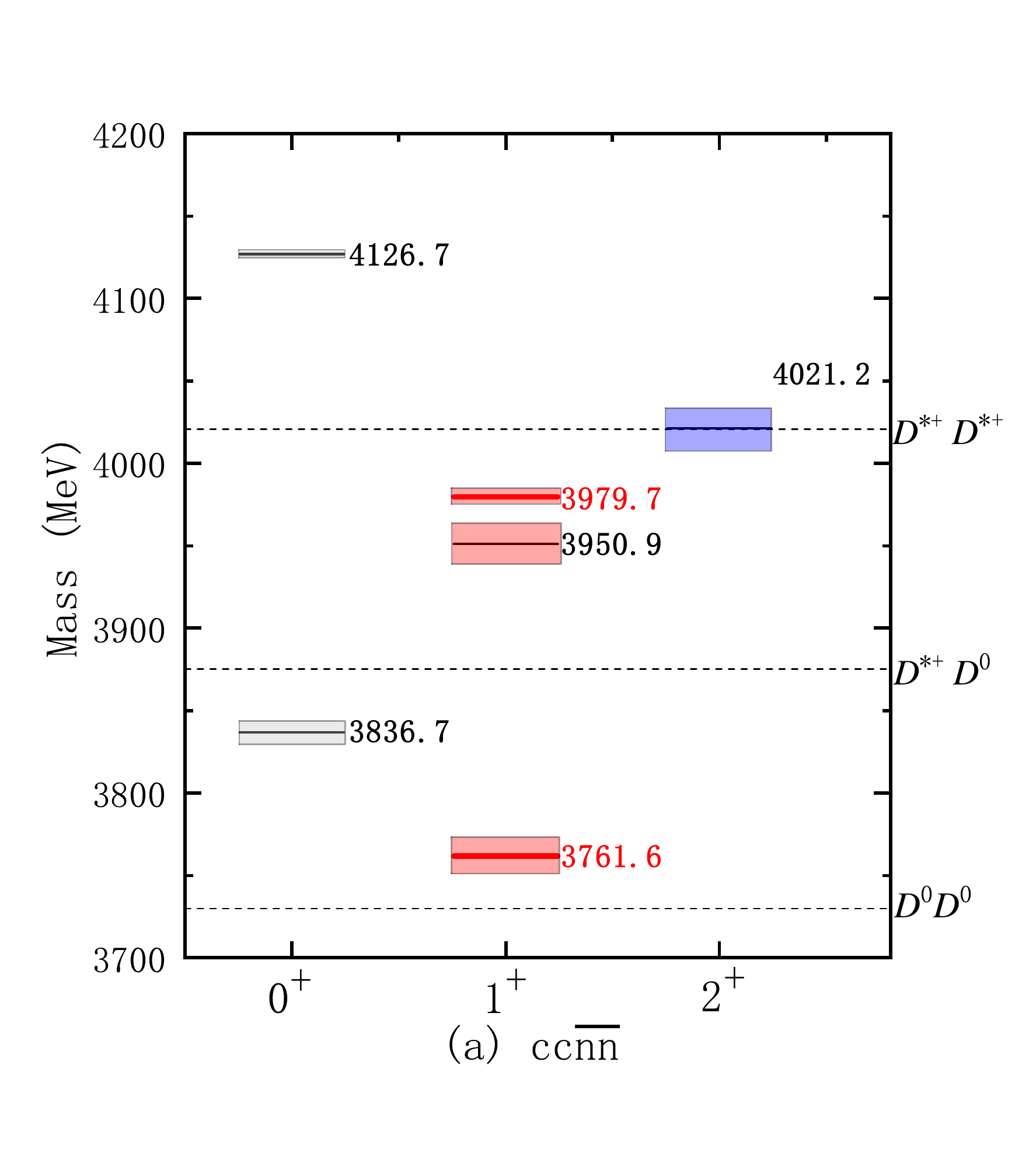}
\end{minipage}
\hspace{-0.85cm}
\begin{minipage}{6.39cm}
    \includegraphics[width=0.96\textwidth,height=0.96\textwidth]{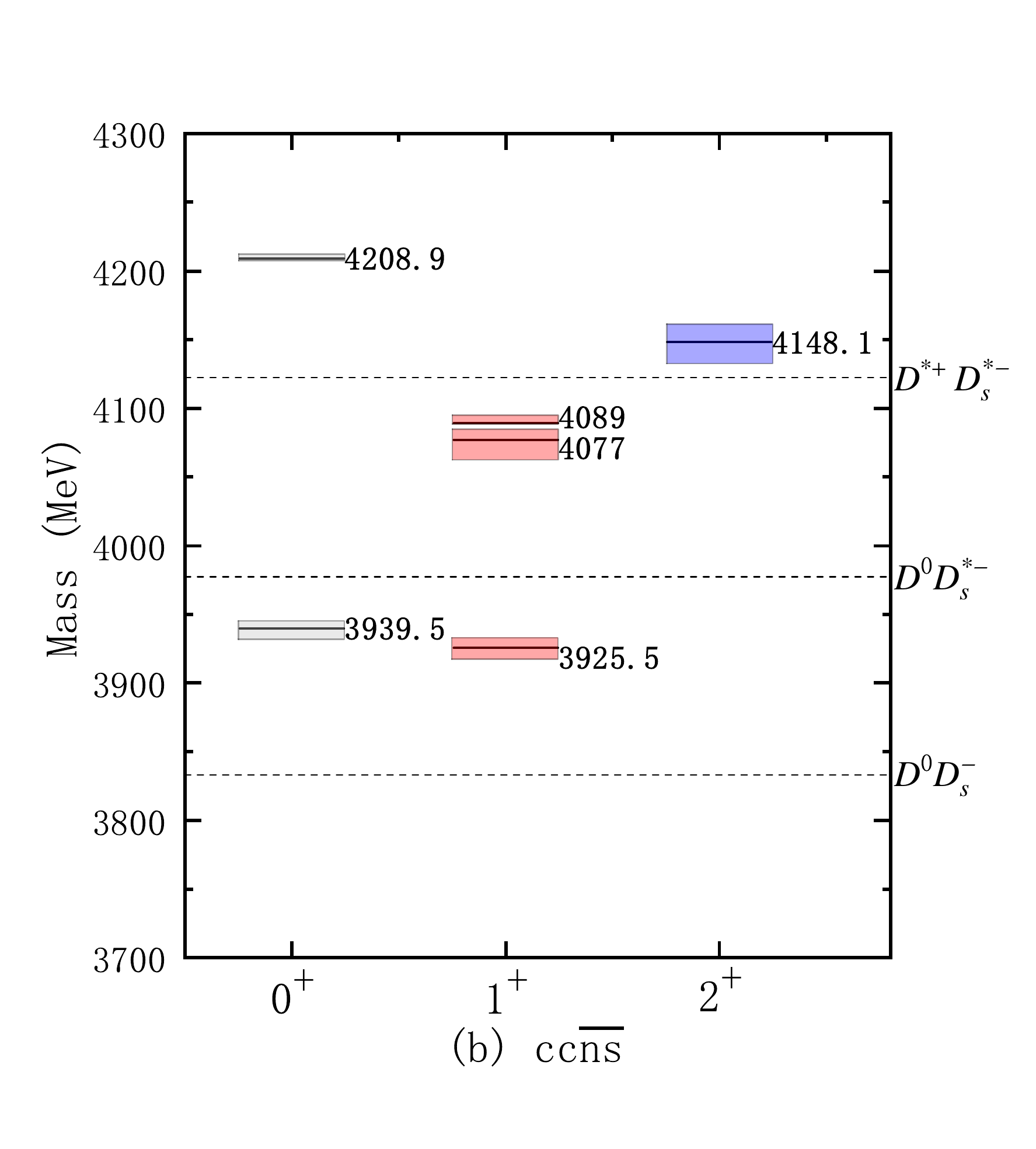}
\end{minipage}
\hspace{-0.85cm}
\begin{minipage}{6.39cm}
    \includegraphics[width=0.96\textwidth,height=0.96\textwidth]{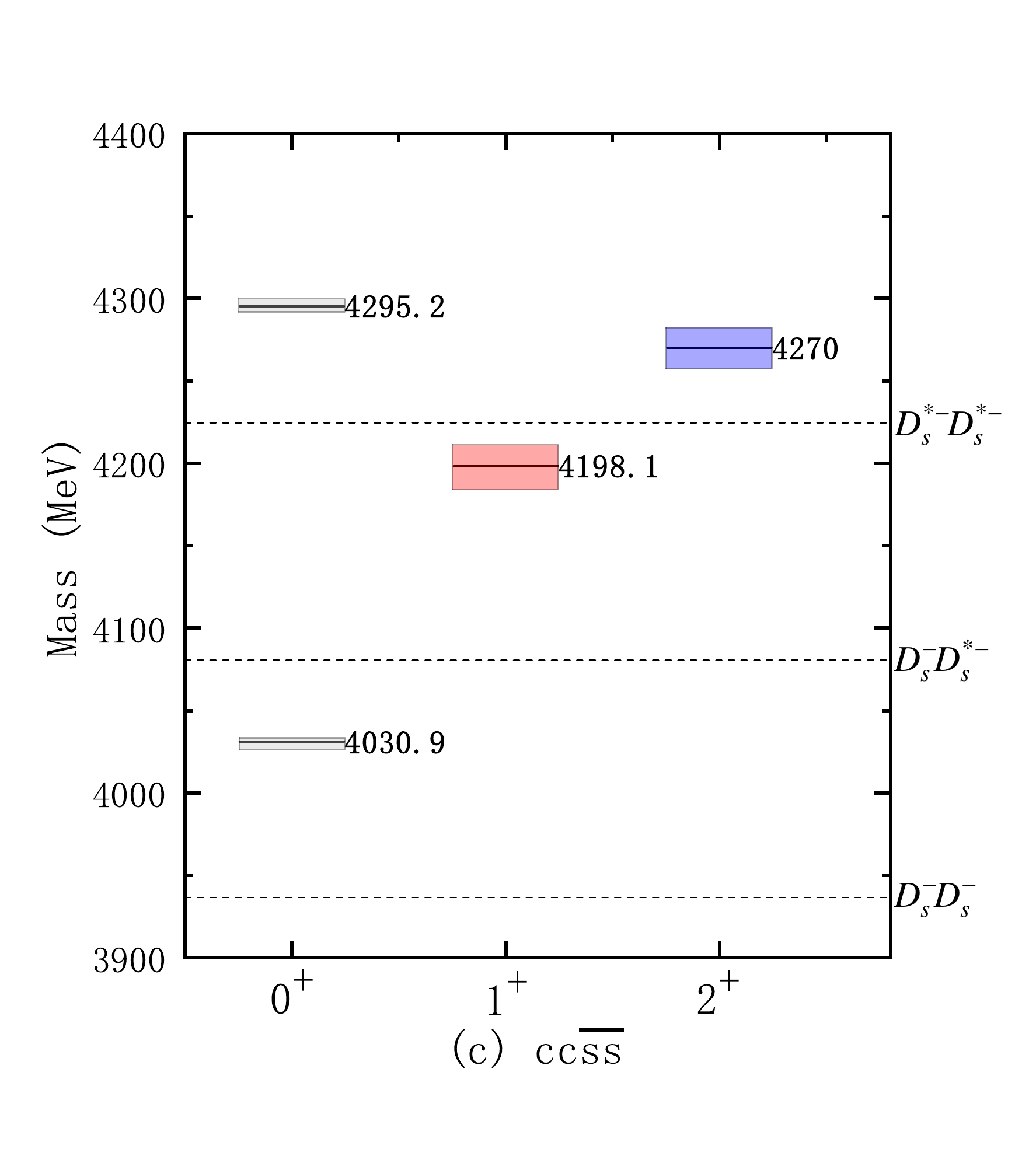}
\end{minipage}

\vspace{-0.5cm}
\caption{\label{f1}
Mass spectra of $S$-wave tetraquark states $cc\bar{n}\bar{n}$, $cc\bar{n}\bar{s}$, and $cc\bar{s}\bar{s}$ with quantum numbers $J^P=0^+$, $1^+$, and $2^+$.
The thin black (thick red) solid lines in (a) denote the isospin triplet (singlet) states.
The bands represent the uncertainties of the masses. The black dashed lines are the corresponding meson-meson thresholds.}
\end{figure*}

\begin{figure*}[htbp]
\centering
\begin{minipage}{6.39cm}
    \includegraphics[width=0.96\textwidth,height=0.96\textwidth]{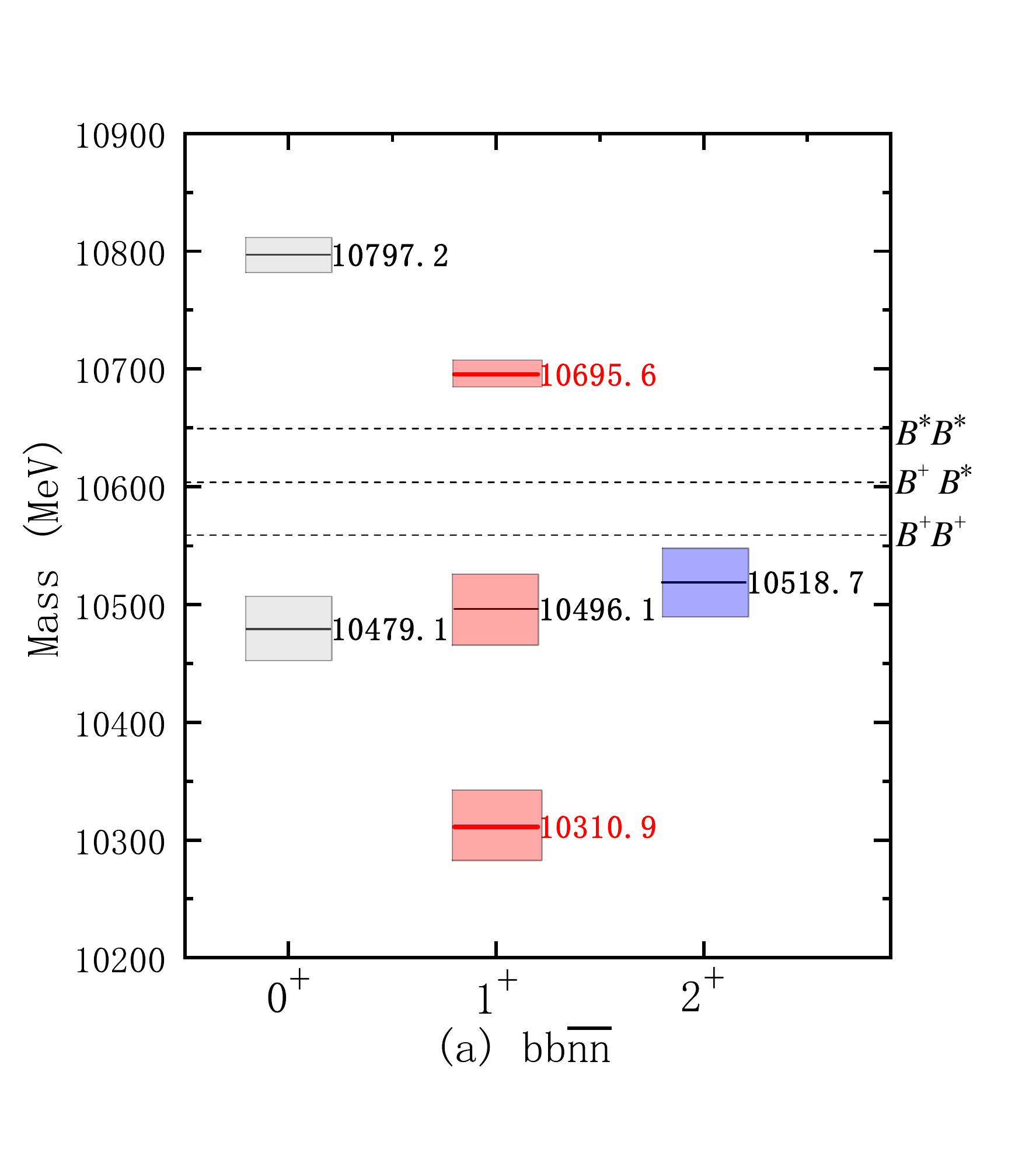}
\end{minipage}
\hspace{-0.85cm}
\begin{minipage}{6.39cm}
    \includegraphics[width=0.96\textwidth,height=0.96\textwidth]{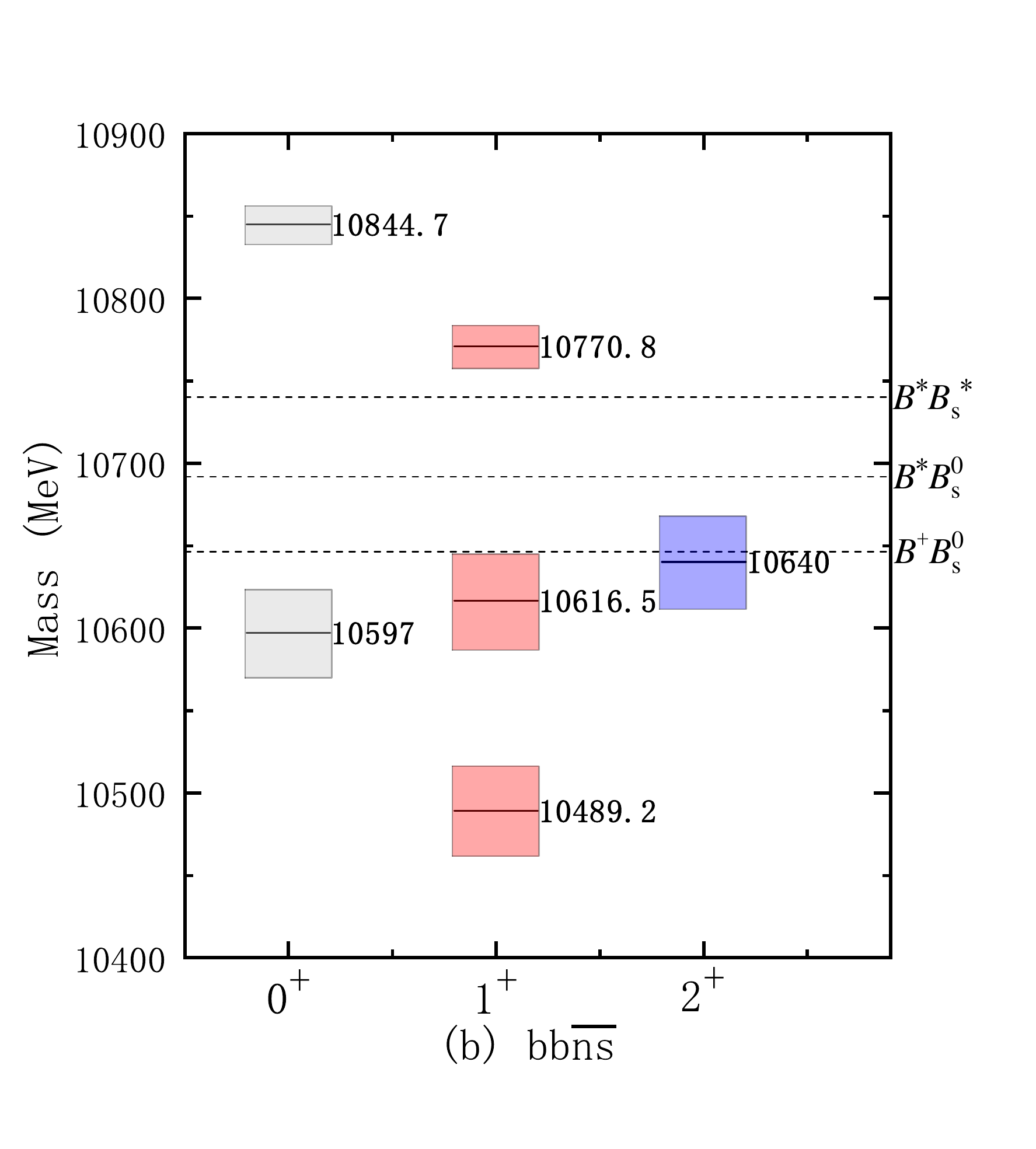}
\end{minipage}
\hspace{-0.85cm}
\begin{minipage}{6.39cm}
    \includegraphics[width=0.96\textwidth,height=0.96\textwidth]{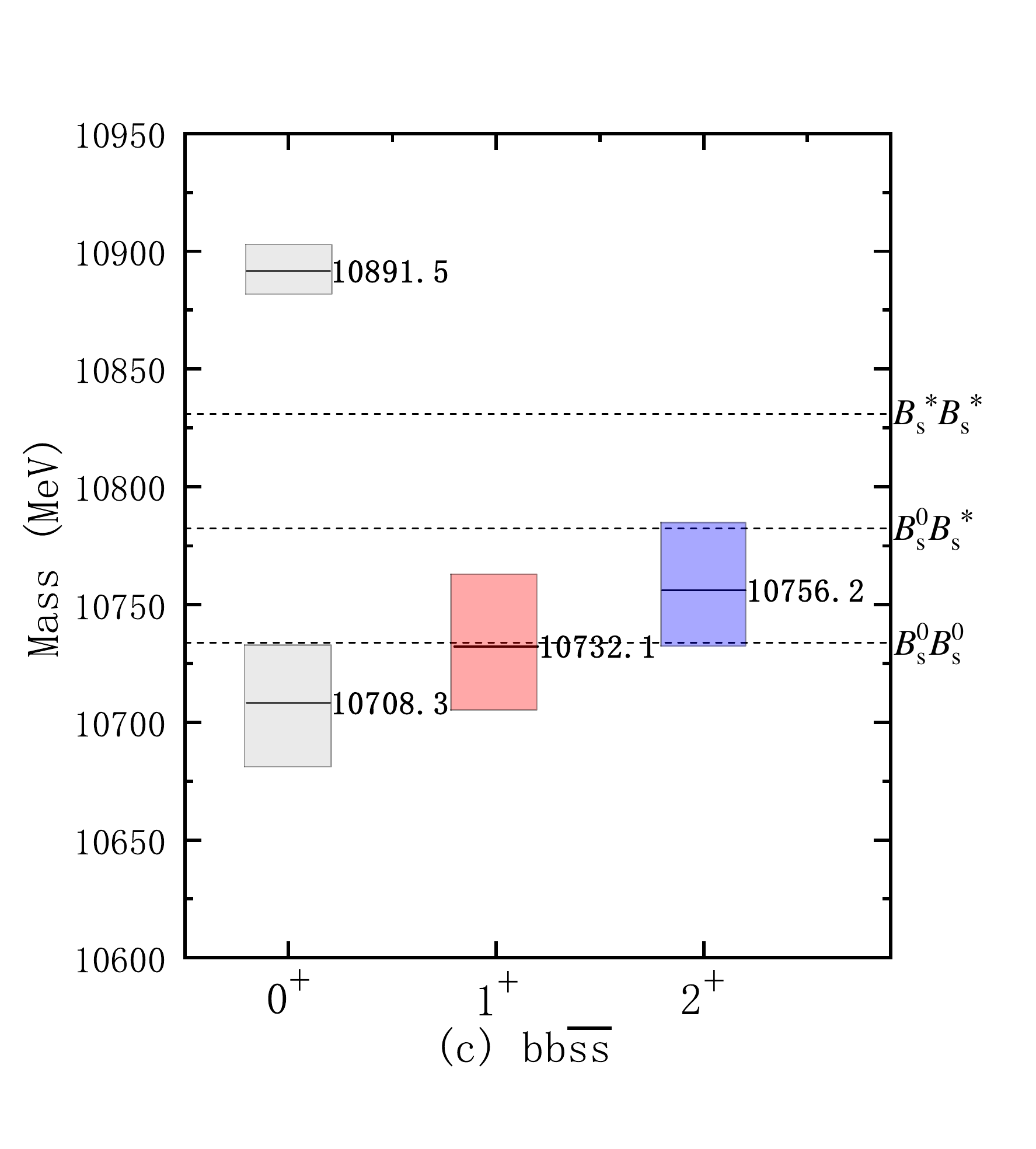}
\end{minipage}

\vspace{-0.5cm}
\caption{\label{f2}
Mass spectra of $S$-wave tetraquark states $bb\bar{n}\bar{n}$, $bb\bar{n}\bar{s}$, and $bb\bar{s}\bar{s}$ with quantum numbers $J^P=0^+$, $1^+$, and $2^+$.
The thin black (thick red) solid lines in (a) denote the isospin triplet (singlet) states.
The bands represent the uncertainties of the masses. The black dashed lines are the corresponding meson-meson thresholds.}
\end{figure*}

\begin{figure*}[htbp]
\centering
\begin{minipage}{6.39cm}
    \includegraphics[width=0.96\textwidth,height=0.96\textwidth]{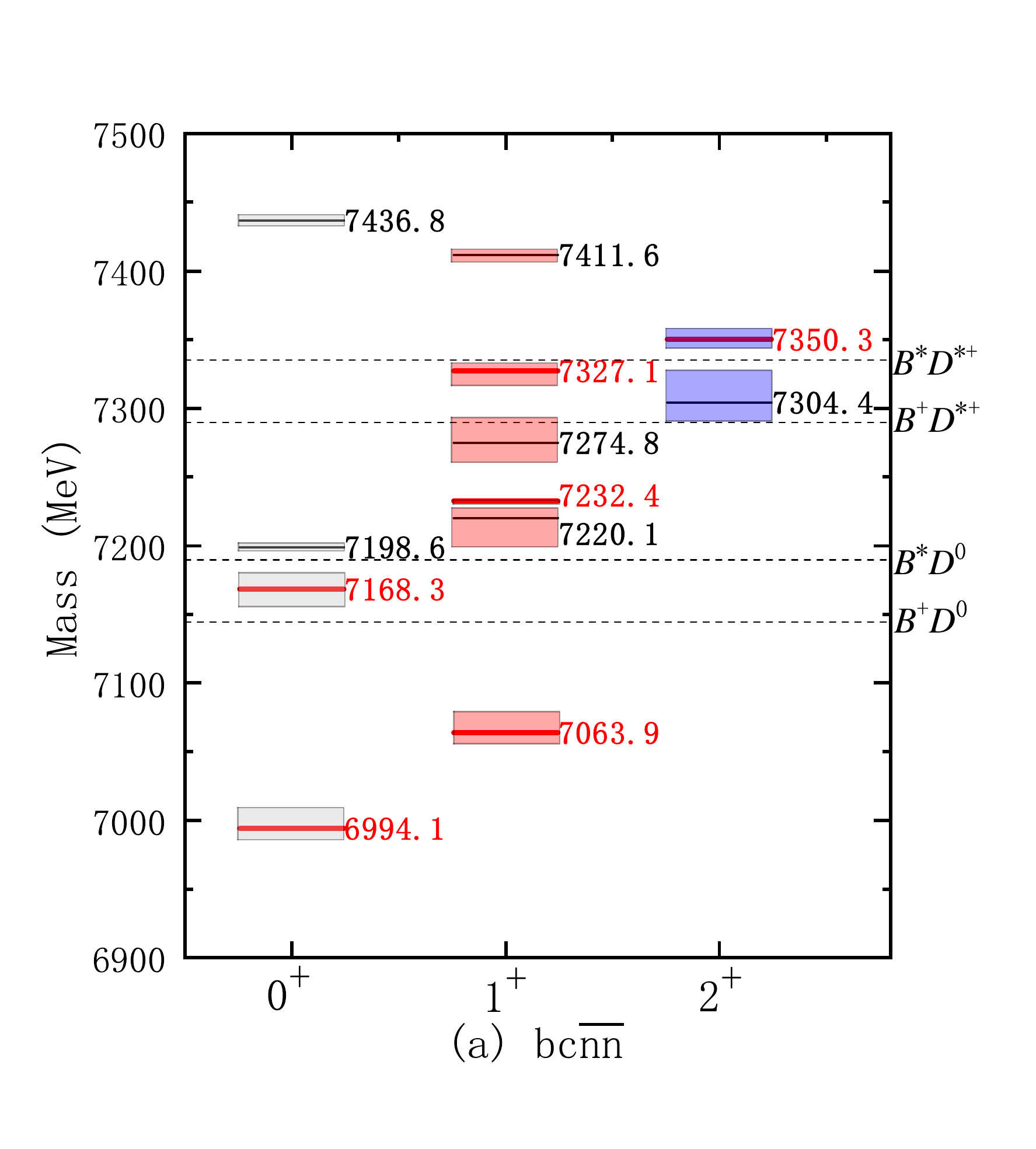}
\end{minipage}
\hspace{-0.85cm}
\begin{minipage}{6.39cm}
    \includegraphics[width=0.96\textwidth,height=0.96\textwidth]{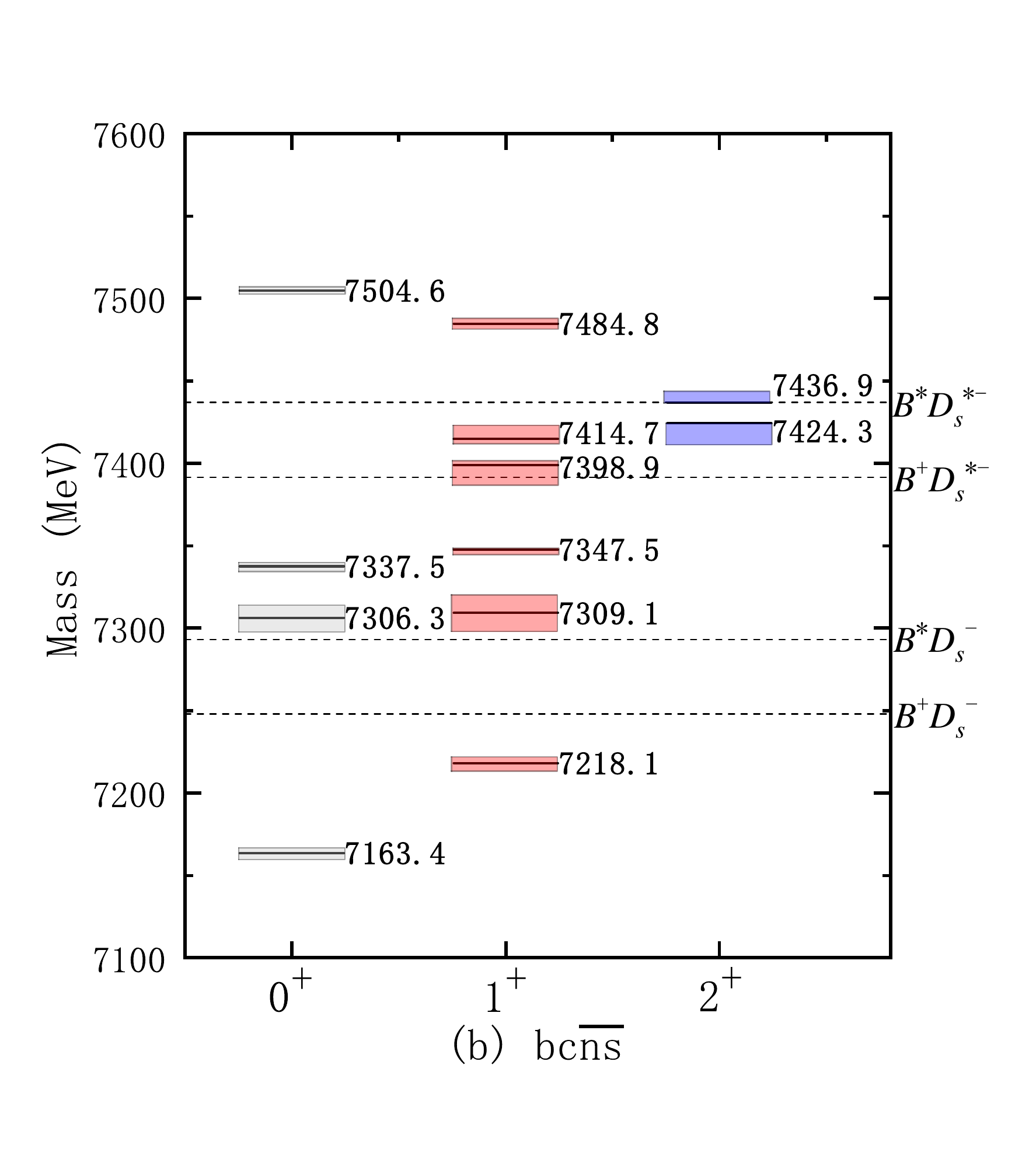}
\end{minipage}
\hspace{-0.85cm}
\begin{minipage}{6.39cm}
    \includegraphics[width=0.96\textwidth,height=0.96\textwidth]{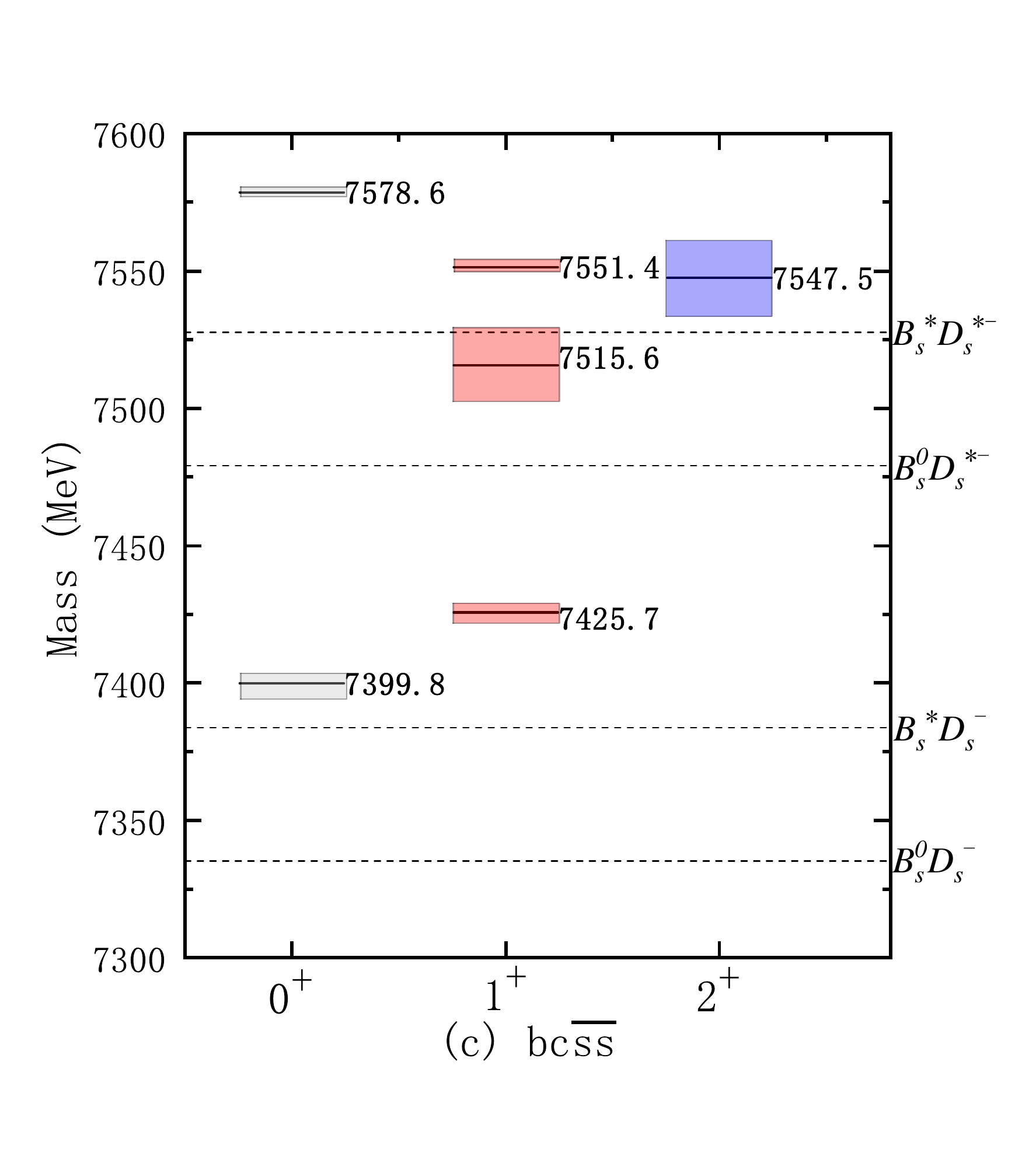}
\end{minipage}

\vspace{-0.5cm}
\caption{\label{f3}
Mass spectra of $S$-wave tetraquark states $bc\bar{n}\bar{n}$, $bc\bar{n}\bar{s}$, and $bc\bar{s}\bar{s}$ with quantum numbers $J^P=0^+$, $1^+$, and $2^+$.
The thin black (thick red) solid lines in (a) denote the isospin triplet (singlet) states.
The bands represent the uncertainties of the masses. The black dashed lines are the corresponding meson-meson thresholds.}
\end{figure*}

With the model parameters listed in Table~\ref{mesonpm} and Table~\ref{baryonpm}, now we can calculate the mass spectra and wave functions of the $S$-wave doubly heavy tetraquark states $QQ\bar{q}\bar{q}$ with quantum numbers $J^P=0^+$, $1^+$, and $2^+$.  The mass spectra of the tetraquark states $cc\bar{q}\bar{q}$, $bb\bar{q}\bar{q}$, $bc\bar{q}\bar{q}$ ($q=u, d, s$) are plotted in Fig.\ref{f1}, Fig.~\ref{f2}, and Fig.~\ref{f3}, respectively. The corresponding meson-meson thresholds are also shown in these plots for comparison. In the calculations, we have taken into account the 
errors of the model parameters $m_{QQ}$ and $v_{QQ}$, which are obtained by covering the mass spectra of doubly and triply heavy baryons predicted by lattice QCD, QCD sum rules, and potential models.  The calculated masses of the tetraquark states $QQ\bar{q}\bar{q}$  therefore acquire error bars as shown in Fig.\ref{f1}, Fig.~\ref{f2}, and Fig.~\ref{f3}. In the following, we focus on the tetraquark state $cc\bar{n}\bar{n}$ and briefly discuss other tetraquark states. We use $n$ to denote the light quarks ($u$ and $d$ quarks) from now on.

The mass spectra of the tetraquark system $cc\bar{n}\bar{n}$ are exhibited in Fig.~\ref{f1}(a). The relevant meson-meson thresholds  $D^0D^0$, $D^{\ast+}D^0$, $D^{\ast+}D^{\ast+}$ are also displayed for comparison. \\
(i) We find two $S$-wave tetraquark states with quantum number $J^P=0^+$.
The lowest state has a mass 3836.7 MeV, which is above the $D^0D^0$  threshold but below the $D^{*+}D^0$ threshold. Thus the decay into the two-meson state $D^0+D^0$ is allowed.
The other state has a larger mass, 4126.7MeV, which allows the decay into the two-meson states  $D^0+D^0$ and $D^{\ast+}+D^{\ast+}$.\\ 
(ii) We find three $S$-wave tetraquark states with quantum number $J^P=1^+$.
While all three states are above the $D^0D^0$ threshold,  the decay into the two-meson state $D^0+D^0$ is only allowed through the $P$-wave process due to the parity (or angular momentum) conservation ($D^0$ is a scalar meson, while the quantum number of the tetraquark state $cc\bar{n}\bar{n}$ is $J^P=1^+$).
The lowest state is an isospin singlet, i.e., an isoscalar tetraquark state. It has a mass 3761.6 MeV and is lower than the $D^{\ast+} D^0$ threshold. Therefore, this state could be a narrow resonance under the strong interactions.
Apart from the value of the mass (which replies on the model parameters), the quantum number ($J^P=1^+$) and the isospin property (isoscalar) are consistent the recently observed narrow exotic tetraquark state,  the doubly charmed tetraquark state $T_{cc}^+$~\cite{LHCb:Polyakov-1,LHCb:Polyakov-2}.
The color-spin wave function, i.e., the amplitudes corresponding to the bases (\ref{1324basis1}) read 
\begin{eqnarray}
(0, -0.930, 0, 0.368, 0, 0).
\end{eqnarray}
In this diquark-anti-diquark configuration, we find that the tetraquark state is dominated by the color-triplet component. It can be transformed to the meson-meson configuration, by using the basis wave functions
\begin{eqnarray}
&&|(Q_1\bar q_3)^1_0\otimes(Q_2\bar{q}_4)^1_1\rangle , \nonumber\\
&&|(Q_1\bar q_3)^1_1\otimes(Q_2\bar{q}_4)^1_0\rangle, \nonumber\\
&&|(Q_1\bar q_3)^1_1\otimes(Q_2\bar{q}_4)^1_1\rangle,\nonumber\\
&&|(Q_1\bar q_3)^8_0\otimes(Q_2\bar{q}_4)^8_1\rangle,\nonumber\\
&&|(Q_1\bar q_3)^8_1\otimes(Q_2\bar{q}_4)^8_0\rangle,\nonumber\\
&&|(Q_1\bar q_3)^8_1\otimes(Q_2\bar{q}_4)^8_1\rangle.
\end{eqnarray}
The corresponding amplitudes are given by 
\begin{eqnarray}
(0.419, -0.419, -0.167, -0.273, 0.273, 0.687).
\end{eqnarray}
The first two coefficients are relatively large,  indicating that the $D^{\ast+}D^0$ is the dominant decay channel for this tetraquark state. In the classical point of view,  the decay into the two meson state $D^{\ast+}+D^0$ is impossible due to the energy conservation. However, since the mass of this state is larger than the total mass of the $D^0D^0\pi^+$ system,  the decay process $X(3761.6)\to D^0+D^0+\pi^+$ is allowed through the quantum off-shell process  $X(3761.6)\to D^{\ast+}+D^0$ and the decay process $D^{\ast+}\to D^0+\pi^+$. Since this state is close to the $D^{\ast+}D^0$ threshold, the decay width could be small. This picture is consistent with the fact that the observed exotic state $T_{cc}^+$ is rather narrow (the width is measured to be a few hundred keV).
The masses of other two partner states are 3950.9 MeV and 3979.7 MeV,  which are both above the $D^{\ast+}D^0$ threshold.\\
(iii) For the quantum number $J^P=2^+$, we find only one tetraquark state. The central value of its mass is 4021.2 MeV, which is very close to the $D^{\ast+}D^{\ast+}$ threshold but well above the $D^{\ast+}D^0$ threshold.

\begin{table}[htbp]
	\caption{\label{tetrastable} Masses and quantum numbers of the predicted stable $S$-wave tetraquark states in the ICMI model.}
	\begin{ruledtabular}
		\begin{tabular}{lcc}
		system	& $IJ^P$ & mass (MeV) \\
			\hline			
			$bb\bar{n}\bar{n}$& $10^+$& 10479.1\\
			                  & $01^+$& 10310.9\\
			                  & $11^+$& 10496.1\\			
			                  & $12^+$& 10518.7\\			
            \hline
            $bc\bar{n}\bar{n}$& $00^+$& 6994.1\\
                              & $01^+$& 7063.9\\
            \hline \hline                  
      system	& $J^P$ & mass (MeV) \\        
            \hline
            $bb\bar{n}\bar{s}$& $0^+$& 10597\\
            & $1^+$& 10489.2\\
            & $1^+$& 10616.5\\					
            \hline
            $bb\bar{s}\bar{s}$& $0^+$& 10708.3\\
            \hline
            $bc\bar{n}\bar{s}$& $0^+$& 7163.4\\
                              & $1^+$& 7218.1\\                    
		\end{tabular}
	\end{ruledtabular}
\end{table}

The mass spectra of the $S$-wave tetraquark states $cc\bar{n}\bar{s}$ and $cc\bar{s}\bar{s}$ are shown in Fig.~\ref{f1}(b) and Fig.~\ref{f1}(c).  The mass spectra of the doubly bottomed tetraquark states ($bb\bar{n}\bar{n}$, $bb\bar{n}\bar{s}$, and $bb\bar{s}\bar{s}$) and the mixed charm-bottom tetraquark states ($bc\bar{n}\bar{n}$, $bc\bar{n}\bar{s}$, and $bc\bar{s}\bar{s}$) can also be computed in the ICMI model. The results are shown in Fig.~\ref{f2}(a)-(c) and Fig.~\ref{f3}(a)-(c). Comparing with the corresponding meson-meson thresholds, we find several candidates for stable tetraquark states. The masses and quantum numbers of these tetraquark states are listed in Table~\ref{tetrastable}. In real world, these tetraquarks could be narrow states that can be observed in future experiments.

\section{summary}
We have performed a systematic study of the $S$-wave doubly heavy tetraquark states $QQ\bar{q}\bar{q}$ ($q=u, d, s$ and $Q=c, b$) in the ICMI model. The parameters in the ICMI model are obtained by fitting the known hadron spectra and are used directly to predict the mass spectra of the doubly heavy tetraquark states. For heavy quarks, the uncertainties of the parameters ($m_{QQ}$ and $v_{QQ}$) are taken into account by covering the doubly and triply heavy baryon masses predicted by lattice QCD, QCD sum rules, and potential models. The mass spectra of the $S$-wave doubly heavy tetraquark states with quantum numbers $J^P=0^+$, $1^+$, and $2^+$ are presented and analyzed. The results indicate that there may exist several exotic bound states in the doubly charmed, doubly bottomed, and mixed charm-bottom sectors.  For the doubly charmed system, we find a stable tetraquark state $cc\bar u\bar d$ below the $D^{\ast+} D^0$ threshold, with quantum number $IJ^P=01^+$. The properties of this state are consistent with the recently observed narrow exotic state,  the doubly charmed tetraquark state $T_{cc}^+$~\cite{LHCb:Polyakov-1,LHCb:Polyakov-2}. Our prediction of other exotic tetraquark states can be examined in future experiments. 
 
{\bf Note Added}: During the preparation of this manuscript, we became aware of the work~\cite{Weng:2021hje}, in which the same doubly heavy tetraquark state was also studied within the chromomagnetic interaction model.

\begin{acknowledgments}
This work is supported by the NSFC under grant Nos. 11775123, 11890712, and 12047535.
\end{acknowledgments}



\end{document}